\documentclass[a4paper]{jpconf}

\usepackage{graphicx}
\usepackage{xcolor}

\newcommand{\cupid}{\ensuremath{\rm CUPID\hbox{-}0}}

\newcommand{\exposure}{9.95~kg$\times$yr}

\newcommand{\startDAQ}{June 2017} 
\newcommand{\stopDAQ}{December 2018} 
\newcommand{\HalfLife}{T_{1/2}^{2\nu} = [8.6~^{+0.2}_{-0.1}] \times 10^{19}~\textrm{yr}}

\newcommand{\xSSD}{255/254}
\newcommand{\xCA}{360/254}

\newcommand{\ZS}{Zn$^{82}$Se}
\newcommand{\se}{$^{82}$Se}

\newcommand{\al}{$\alpha$}

\newcommand{\QBB}{$\mathrm{Q}_{\beta\beta}$}

\newcommand{\vless}{0$\nu\beta\beta$}
\newcommand{\vv}{2$\nu\beta\beta$}

\begin{document}
\title{Results on \se~\vv~with CUPID-0 Phase I}

\author{L~Pagnanini$^{1,2*}$, O~Azzolini$^{2}$, J~W~Beeman$^{3}$, F~Bellini$^{5,6}$, M~Beretta$^{1,2}$, M~Biassoni$^{2}$, C~Brofferio$^{1,2}$, C~Bucci$^{7}$, S~Capelli$^{1,2}$, L~Cardani$^{6}$, P~Carniti$^{1,2}$, N~Casali$^{6}$, D~Chiesa$^{1,2}$, M~Clemenza$^{1,2}$, O~Cremonesi$^{2}$, A~Cruciani$^{6}$, I~Dafinei$^{5,6}$, S~Di~Domizio$^{8,9,}$, F~Ferroni$^{10,6}$, L~Gironi$^{1,2}$, A~Giuliani$^{11,12}$, P~Gorla$^{7}$, C~Gotti$^{2}$, G~Keppel$^{3}$, M~Martinez$^{5,6}$, S~Nagorny$^{7,10}$, M~Nastasi$^{1,2}$, S~Nisi$^{7}$, C~Nones$^{13}$, D~Orlandi$^{7}$, M~Pallavicini$^{8,9}$, L~Pattavina$^{7}$ M~Pavan$^{1,2}$, G~Pessina$^{2}$, V~Pettinacci$^{5,6}$, S~Pirro$^{7}$ S~Pozzi$^{1,2}$, E~Previtali$^{1,2}$, A~Puiu$^{1,2}$,C~Rusconi$^{7,14}$, K~Sch\"affner$^{7,10}$, C~Tomei$^{6}$, M~Vignati$^{6}$, and A~S~Zolotarova$^{13}$}

\address{$^{1}$Dipartimento di Fisica, Universit\'a di Milano Bicocca, I-20126 Milano, Italy\\
$^{2}$INFN Sezione di Milano - Bicocca, I-20126 Milano, Italy\\
$^{3}$INFN Laboratori Nazionali di Legnaro, I-35020 Legnaro (Pd), Italy\\
$^{4}$Lawrence Berkeley National Laboratory, Berkeley, California 94720, USA\\
$^{5}$Dipartimento di Fisica, Sapienza Universit\`a di Roma, P.le Aldo Moro 2, 00185, Roma, Italy\\
$^{6}$INFN, Sezione di Roma, P.le Aldo Moro 2, 00185, Roma, Italy\\
$^{7}$INFN Laboratori Nazionali del Gran Sasso, I-67100 Assergi (AQ), Italy\\
$^{8}$Dipartimento di Fisica, Universit\`a di Genova, I-16146 Genova, Italy\\
$^{9}$INFN Sezione di Genova, I-16146 Genova, Italy\\
$^{10}$Gran Sasso Science Institute, 67100, L'Aquila, Italy\\
$^{11}$CSNSM, Univ. Paris-Sud, CNRS/IN2P3, Universit\'e Paris-Saclay, 91405 Orsay, France\\
$^{12}$ DISAT, Universit\`a dell'Insubria, 22100 Como, Italy\\
$^{13}$ IRFU, CEA, Universit\`e Paris-Saclay, F-91191 Gif-sur-Yvette, France\\
$^{14}$ Department of Physics and Astronomy, University of South Carolina, Columbia, South Carolina 29208, USA}
\ead{lorenzo.pagnanini@gssi.it}

\begin{abstract}
The nucleus is an extraordinarily complex object where fundamental forces are at work. The solution of this many-body problem has challenged physicists for decades: several models with complementary virtues and flaws have been adopted, none of which has a universal predictive capability. Double beta decay is a second order weak nuclear decay whose precise measurement might steer fundamental improvements in nuclear theory. Its knowledge paves the way to a much better understanding of many body nuclear dynamics and clarifies, in particular, the role of multiparticle states. This is a useful input to a complete understanding of the dynamics of neutrino-less double beta decay, the chief physical process whose discovery may shed light to matter-antimatter asymmetry of the universe and unveil the true nature of neutrinos.
Here, we report the study of \vv-decay in \se~ with the CUPID-0 detector, an array of ZnSe crystals maintained at a temperature close to `absolute zero' in an ultralow background environment. 
Thanks to the unprecedented accuracy in the measurement of the two electrons spectrum, we prove that the decay is dominated by a single intermediate state.
We obtain also the most precise value for the $^{82}$Se~\vv-decay half-life of $\HalfLife$.
\end{abstract}

\section{Introduction}
\begin{figure}
\centering 
\includegraphics[width=11cm]{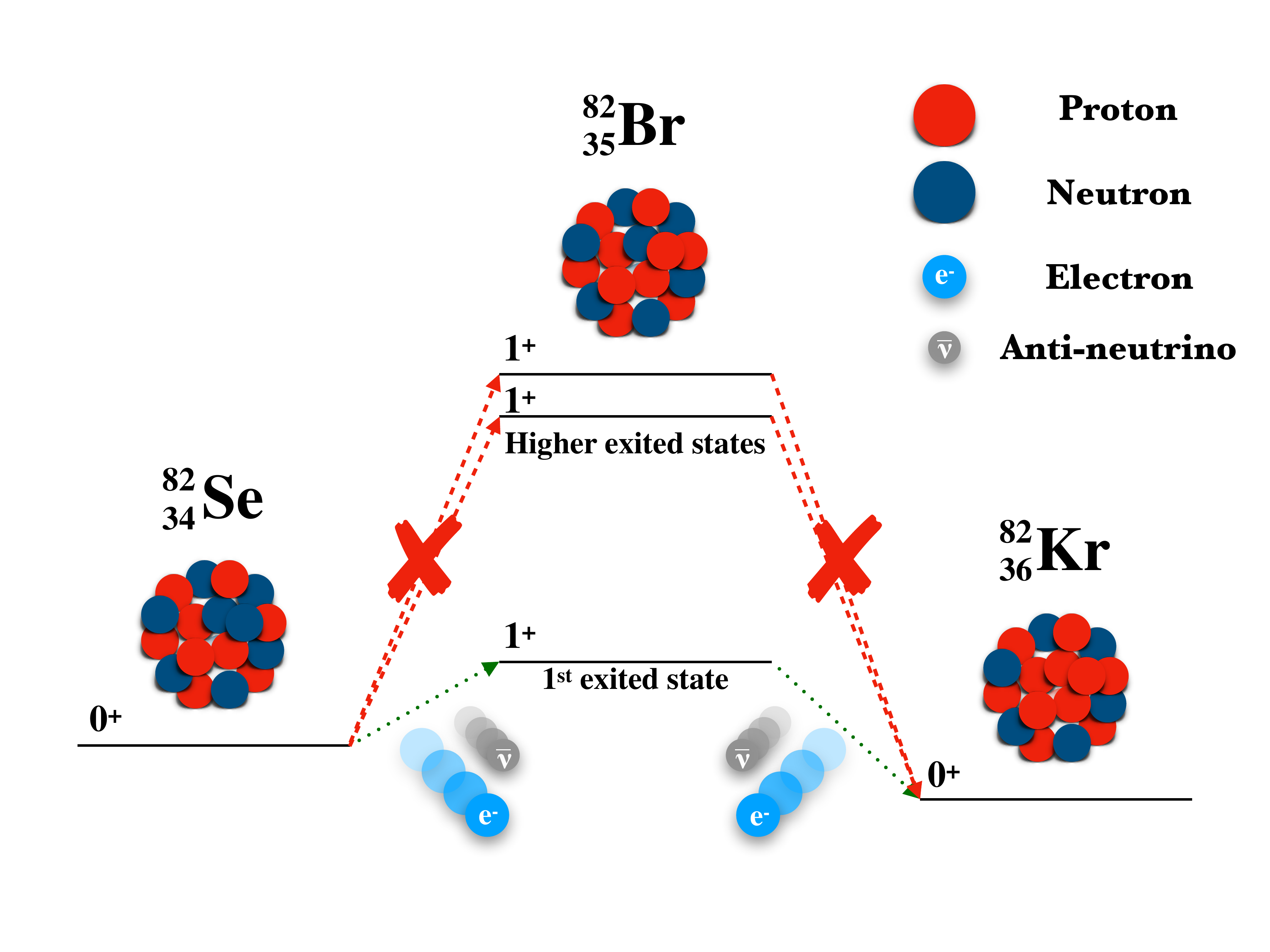}
\caption{{Schematic of the two-neutrino double beta decay.} The transition between the initial state (\se) and the final one ($^{82}$Kr) is represented as two consecutive $\beta$-decays that have $^{82}$Br as an intermediate step. The decay can occur through two different paths involving practically only the first 1$^{+}$ exited state of $^{82}$Br (green dotted arrows) or also high-lying levels (red dashed arrows).}
\label{fig:process}
\end{figure}

In 1935, one year after the Fermi formulation of the theory of $\beta$-decay, M. Goeppert-Mayer speculated in Physical Review \cite{GoeppertMayer:1935qp} about the existence in nature of multiple stable isobars: \emph{`Why in geologic time they have not all be transformed into the most stable [lightest] isobar by consecutive $\beta$-decays?'}.
A look at the chart of the nuclides shows that we have several $(A,Z)$ nuclei more strictly bounded (i.e., lighter) than their $(A, Z+1)$ isobars but less bounded (i.e., heavier) than their $(A, Z+2)$ isobars. 
The consequence is that in many cases, the $(A,Z)$ nucleus cannot $\beta$-decay and appears stable. 
M. Goeppert-Mayer suggested a viable, though rare, decay mechanism leading even-even nuclei to the lightest isobar: two neutrino double-$\beta$ (\vv) decay or symbolically
$(A,Z) \rightarrow (A,Z+2) + 2e^-+2\overline{\nu}_e$. This is the rarest weak decay ever detected, which
today is observed in eleven isotopes with halflives ranging from 10$^{18}$ to 10$^{24}$ yr  \cite{Barabash:2019nnr}.
In the modern language of fundamental physics, \vv-decay is a second-order weak transition. The difference in the mass of the two isobars, \QBB, accounts for the total kinetic energy shared by the outgoing particles. 
The decay is modeled as a two-step process in which the intermediate nucleus $(A, Z+1)$ is accessed through a `virtual' transition (Fig. \ref{fig:process}), i.e. violating energy conservation within the limits of the uncertainty principle. 

The decay probability is obtained by adding the contributions of each accessible state of the intermediate nucleus, and the condition of spin-parity changes restricts these states to the levels with J$^P$=1$^+$. Nuclear theory is not yet able to say if the dominant term comes from the lowest energy 1$^{+}$ state or if higher states are also involved. The two cases, labeled as single state dominant (SSD) and higher state dominant (HSD), result in slightly different predictions for the differential distributions of the two electrons' energies and therefore can be disentangled by an accurate measurement of the \vv-decay spectrum.
Here we report the measurement of the \se~\vv-decay half-life and an analysis of its spectral shape that proves the SSD character of the nuclear transition. This hypothesis is discouraged by nuclear theories for \se~and supported for $^{96}$Zr, $^{100}$Mo, and $^{116}$Cd \cite{Moreno:2008dz}, since the corresponding odd-odd nuclei have a low-lying (1g$_{9/2}$ 1g$_{7/2}$) 1$^{+}$ level. On the contrary, the low-lying 1$^{+}$ state of $^{82}$Br is (2p$_{1/2}$ 2p$_{3/2}$), and is not the ground state (which is 5$^{-}$).
Previous experimental searches for \se~were performed only by the NEMO-3 experiment  \cite{Arnold:2004xq}, which investigated several \vv-emitters (e.g.~$^{48}$Ca, $^{82}$Se, $^{100}$Mo, $^{116}$Cd). Despite the excellent results recently obtained for $^{100}$Mo  \cite{NEMO-3:2019gwo}, this kind of search for \se~was affected by systematic uncertainties that prevent a clear discovery statement on the nuclear process details \cite{Arnold:2018tmo}.

With the model identified, we can extract the experimental value of the nuclear matrix element that can be used to test nuclear models \cite{Barea:2013bz,Suhonen:2017krv}.
These results mark a significant step toward nuclear model comprehension, the weakness of which  has a high impact on the other double $\beta$-decay channel in which no neutrinos are emitted, $(A,Z) \rightarrow (A,Z+2) + 2e^-$.  Called neutrinoless double-$\beta$ (\vless) decay  \cite{DellOro:2016tmg}, this transition violates lepton number. This case is never observed in physics since lepton production generally occurs in particle-antiparticle pairs. A proof of its existence could explain the long-standing problem of our universe being dominated by matter. The significance of the result explains the number of large experiments dedicated to a hunt for \vless~\cite{Dolinski:2019nrj}.

\section{CUPID-0 experiment}
CUPID-0 is one of these experiments and its mission is twofold: search for $^{82}$Se \vless-decay (\QBB $=2997.9\pm0.3$)  \cite{Lincoln:2012fq}  and demonstrate the excellence of the `scintillating bolometer' technology at the base of the upcoming ton-scale CUPID project. The experiment is located underground at the Laboratori Nazionali del Gran Sasso (Italy) where approximately 1400~m of rock overburden suppresses the cosmic ray flux \cite{Ambrosio:1995cx}. Twenty centimeters of lead shield the cryostat from the mountain radioactivity while a shield, cast from ancient Roman lead  \cite{Pattavina:2019pxw}, surrounds the very core of the experiment: an array of 26 ZnSe scintillating crystals cooled at 10 mK. In particular, 24 crystals are grown from \se~enriched materials, while the remaining two are natural.
Each ZnSe is simultaneously a source and detector of \se~decays. A thermometer and a couple of light detectors record the heat and light signals produced by particles interacting in the crystal. The heat signal provides an accurate measurement of energy, while the ZnSe scintillation signal is used for particle identification. The data we present here were collected between \startDAQ~and \stopDAQ~with an active mass of 8.74 kg of~\ZS~and a \exposure~ \ZS~exposure. The analysis flow starts with the evaluation of amplitude and shape of both heat and light signals, using a matched-filter algorithm. The energy scale and resolution of the heat signal are determined in calibration runs using two $^{232}$Th weak sources. The particle identification capability is established comparing the shape of the corresponding light signals .
A detailed description of the CUPID-0 detector and activity is reported in Refs.~ \cite{Arnaboldi:2017aek} (electronics),  \cite{Dafinei:2017xpc} (crystals production),  \cite{Azzolini:2018tum} (construction and operation),  \cite{DiDomizio:2018ldc} (data acquisition),  \cite{Azzolini:2018yye,Beretta:2019bmm} (analysis techniques), and  \cite{Azzolini:2018dyb,Azzolini:2018oph,Azzolini:2019tta,Azzolini:2019yib,Azzolini:2019nmi} (results).

For the \vv~analysis we select events that satisfy the signature of \vv-decay \cite{Azzolini:2019yib}, and we project the energy spectrum of the surviving events. We reject \al~particles with energies above 2~MeV (below this value particle identification is poor) and events in which more than one crystal is involved. The latter cut removes a tiny fraction of \vv~events since the two electrons are generally stopped inside the same crystal where the \se~decay occurs. Conversely, it removes a significant fraction of background induced events such as $\gamma$'s from radioactive contamination or $\mu$'s from cosmic rays. 
The probability that a \vv~signal survives all the cuts is constant above 150 keV and equal to $\varepsilon_{C}=(95.7 \pm 0.5)\%$ (details in \cite{Azzolini:2019nmi}). 
\section{Results on \vv~of \se}
The spectrum of accepted events is dominated by \vv-decay. However, a residual background is expected by \al's (below 2~MeV) and by $\beta/\gamma$'s that interact only in one ZnSe crystal. 
We study this spectrum using the CUPID-0 background model \cite{Azzolini:2019nmi}, in which Monte Carlo simulations are used to explain experimental data.
We produce simulated data for each background source (including \vv-decay) using a code that generates and propagates the particles in the \cupid~geometry. 
Additionally, we reproduce the specific features of the detector as energy resolution, threshold, and \al~particle identification.
The measured background is reconstructed as a linear combination of these sources using a fit procedure that computes individual source activity. 
The fit uses the spectrum of events surviving the \vv~cuts, but in order to reduce the correlation among different background sources, it also exploits the spectrum of \al~events and the spectra in which more than one crystal triggered the event.

We run the fit twice using the \vv-decay spectral shape predicted by one of the two mechanisms, i.e.,~SSD or HSD, and simulated in the \cupid~detector. 

\begin{table}
\caption{Number of events reconstructed by the fit as \vv~Signal (S) and Background (B) from the fit threshold (700 keV) up to the \se~\vv~endpoint (3 MeV). 
The Signal-to-Background ratio calculated from these numbers could be misleading, since the normalization of the \vv~spectrum is driven by the region between 1.6  and 2.5 MeV. For this reason we report also the values obtained in this range.}
\begin{center}
\begin{tabular}{lrr}
\hline
&  1.6 - 2.5 & 0.7 - 3.0\\
& [MeV] & [MeV]\\			
\hline
Signal&	61866 &	246333\\
Background&	3674 &56964\\
S/B& 16.84 & 4.32\\
\hline
\end{tabular}
\end{center}
\label{Tab:SN}
\end{table}

In the SSD scenario, the fit reproduces the experimental spectrum very effectively, with a global $\chi^{2}$/ndf = \xSSD, while for the HSD mechanism the fit quality is quite worse ($\chi^{2}$/ndf = \xCA).
A detailed comparison of the two models can be found in Ref.  \cite{Azzolini:2019yib}. We show in Tab.~\ref{Tab:SN} the number of events reconstructed by the fit as \vv~Signal (S) and Background (B) in the SSD fit, which is assumed as reference.
The normalization of the \vv~spectrum is driven by the region between 1.6  and 2.5 MeV, where  \vv is the dominant component  \cite{Azzolini:2019yib}.
We investigated the impact of binning, threshold, energy scale uncertainty, background model components, and data selection obtaining a global effect smaller than 2 \% on the reconstructed \vv~signal. 
More details can be found in Ref.  \cite{Azzolini:2019yib}. By converting the experimental value of the \vv~activity and its uncertainty in \se~\vv~half-life, we obtain $\HalfLife$, whose uncertainty is three-times lower than the previous NEMO-3 result \cite{Arnold:2018tmo}.

The theoretical expression for the \vv-decay half-live is:
\begin{equation}
\label{half_To_ga}
(T^{2\nu}_{1/2})^{-1} = G_{2\nu} \cdot (g_{A}^{eff})^4 \cdot \mathcal{M}_{2\nu}^{2},
\end{equation}
where $G_{2\nu}$ is the phase-space factor, $\mathcal{M}_{2\nu}$ is the nuclear matrix element, and $g_{A}^{eff}$ is the effective axial coupling constant that in nuclear models is used to account for the discrepancies between data and theoretical predictions.
By calculating $G^{SSD}_{2\nu}=1.996\times10^{-18}$ under the SSD hypothesis \cite{Kotila:2012zza}, we can use the experimental result for \vv-decay half-life to evaluate the effective nuclear matrix element:
\begin{equation}
\mathcal{M}^{eff}_{2\nu} =
(g_A^{eff})^2 \cdot \mathcal{M}_{2\nu} = 0.0762~_{-~0.0006}^{+~0.0005}
\label{Meff}
\end{equation}
where the uncertainty includes both statistical and systematic uncertainty, summed in quadrature. 

Since the \vv~of \se~has been not expected to be single-state dominated, the NME calculated up to now in the framework of different models (IBM  \cite{Barea:2015kwa}, ISM  \cite{Menendez:2009xa} and QRPA \cite{Simkovic:2018hiq}) can not be compared with our result. This will be a useful benchmark for the nuclear models, when the NMEs calculated under the SSD hypothesis will be available.

In summary, we have performed the most precise measurement of \se~\vv~half-life. Moreover, we have established that the \vv-decay of \se~is single state dominated, ruling out the hypothesis that the higher states of the intermediate nucleus participate in this nuclear transition. Our result in Eq.~\ref{Meff} can then be used to test nuclear structure models and extract the associated values of $g_A$ in \vv-decay.

The result presented here is based on a robust model of the CUPID-0 background \cite{Azzolini:2019nmi} and are achieved using ultra-pure scintillating bolometers. 
The wide span of physics results obtained, despite the small exposure, proves once more the potential of this technique, setting an important milestone for the next-generation CUPID experiment.

\section{CUPID-0 Phase II}
\begin{figure}
\centering 
\includegraphics[width=10.5cm]{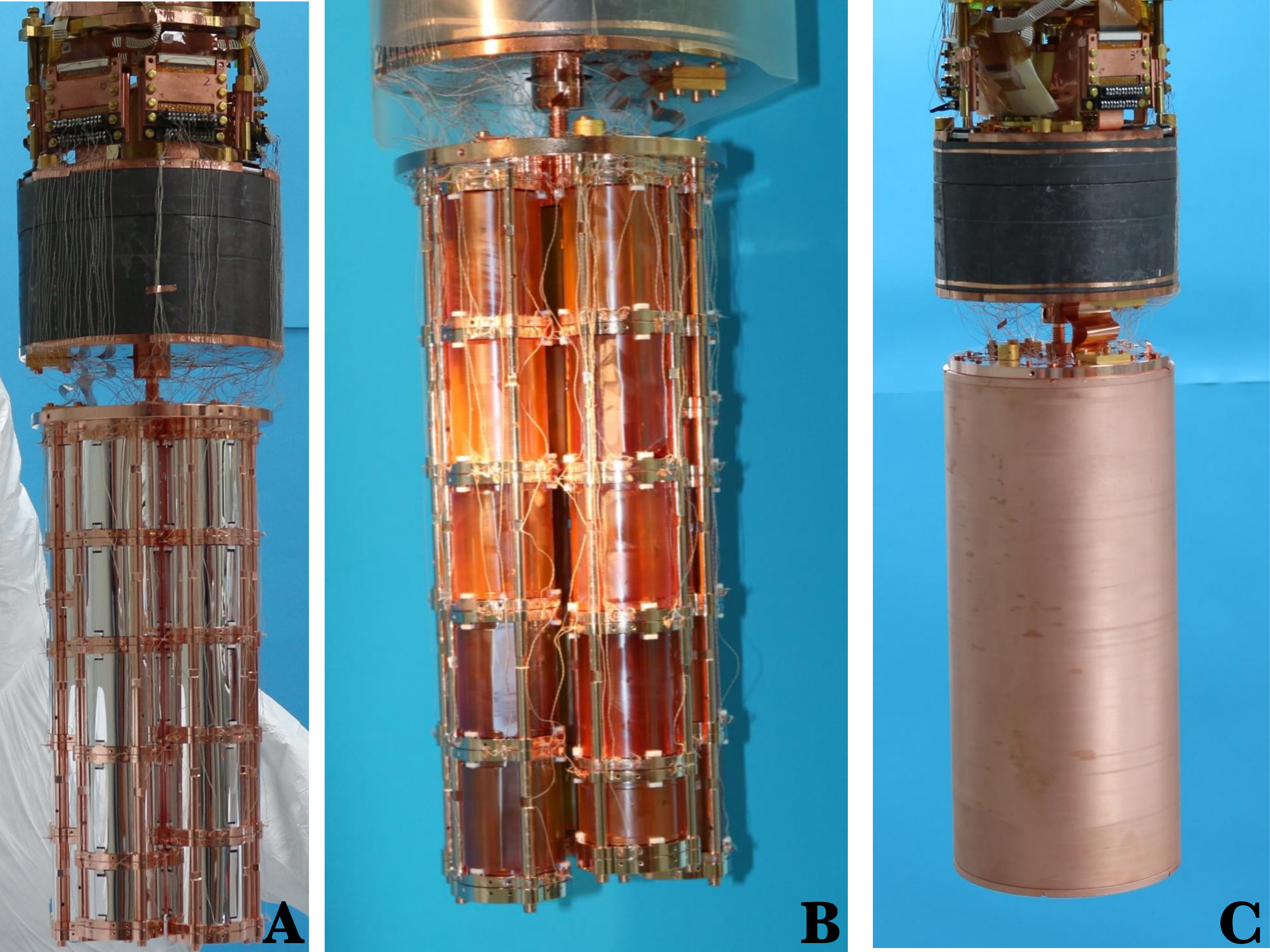}
\caption{Pictures of the CUPID-0 experiment. In the Phase I experimental setup (A) each crystal is laterally surrounded by a reflecting foil, which is removed for the Phase II (B). We add also a 10 mK copper thermal shield (C).}
\label{fig:phaseII}
\end{figure}

After a technical stop from December 2018 to May 2019, CUPID-0 is running since June 2019 with an upgraded detector configuration (Phase II), the design of which has been driven by the background model results.
The main goal of Phase II is to validate the CUPID-0 background model and clarify the origin of the counting rate in the region of interest.
Since a large fraction of our background is expected to be due to muons, we have installed a muon veto based on plastic scintillators.
Moreover, we have removed the reflecting foils (Fig.~\ref{fig:phaseII} B) with the purpose of performing a more detailed study of the crystal surface contamination. 
Indeed, the reflecting foil has prevented to exploit the $\alpha$ double-hit events, which are very useful to estimate the activity of surface $\alpha$-contaminations and to understand their depth profile  \cite{Alduino:2016vtd}.
Finally, we have installed an additional 1 cm thick copper shield (Fig.~\ref{fig:phaseII} C) against the thermal radiation of the 50 mK stage, that was previously absorbed by the reflector.
Such copper shield will reduce also the measured intensity of the $\gamma$-lines produced by the cryostat contamination.
In particular, the $^{232}$Th contaminations produce (in cascade) the 2615 keV and the 583 keV $\gamma$-lines, whose sum falls in the ROI.

\section*{References}
\bibliography{main} 
\bibliographystyle{ieeetr}

\end{document}